%% file: main.tex
\documentclass{sig-alternate} 
\usepackage{mathptmx} 

\newcommand{\ignore}[1]{}
\usepackage{fancyhdr}
\usepackage[normalem]{ulem}
\usepackage[hyphens]{url}
\usepackage{microtype}
\usepackage{multirow}

\usepackage{threeparttable, tablefootnote}

\input{my_definitions}

\usepackage[bookmarks=true,breaklinks=true,letterpaper=true,colorlinks,linkcolor=black,citecolor=blue,urlcolor=black]{hyperref}

\title{
  Recomputation Enabled Efficient Checkpointing
}
\numberofauthors{2}
\author{
	\alignauthor Ismail Akturk\\
	 \affaddr{University of Missouri, Columbia}\\
	 \email{akturki@missouri.edu}
	\alignauthor Ulya R. Karpuzcu\\
	 \affaddr{University of Minnesota, Twin Cities}\\
	 \email{ukarpuzc@umn.edu}	
}

\begin{document}
\maketitle
\pagestyle{plain}

\begin{abstract}
\input{abs}

\end{abstract}

\section{Introduction}
\label{sec:checkpoint_intro}
\input{checkpoint_intro}

\section{Background}
\label{sec:back}
\input{back}

\section{\arch\ Basics}
\label{sec:checkpoint_design}
\input{checkpoint_design}

\section{Evaluation Setup}
\label{sec:checkpoint_setup}
\input{checkpoint_setup}

\section{Evaluation}
\label{sec:checkpoint_eval}
\input{checkpoint_eval}

\section{Related Work}
\label{sec:checkpoint_related}
\input{checkpoint_related}

\section{Conclusion}
\label{sec:conc}
\input{conc}

\bibliographystyle{ieeetr}
\bibliography{references}

\end{document}

%% file: my_definitions.tex
\newcommand{\arch}{AmnesiCHK}

\newcommand{\rs}{{\em RSlice}}

\newcommand{\nockpt}{{\it No}$_{Ckpt}$}
\newcommand{\nfnr}{{\it Ckpt}$_{NE}$}
\newcommand{\wfnr}{{\it Ckpt}$_E$}
\newcommand{\nfwr}{{\it Amn}$_{NE}$}
\newcommand{\wfwr}{{\it Amn}$_E$}
\newcommand{\nfnrl}{{\it Ckpt}$_{NE,Loc}$}
\newcommand{\wfnrl}{{\it Ckpt}$_{E,Loc}$}
\newcommand{\nfwrl}{{\it Amn}$_{NE,Loc}$}
\newcommand{\wfwrl}{{\it Amn}$_{E,Loc}$}

\newcommand{\shrink}{Eliminate vertical white-space}
\newcommand{\vshrink}[1]{
  \ifdefined\shrink 
	\vspace{-#1cm}
  \else
	\vspace{0cm}
  \fi
}


%% file: abs.tex
\noindent 
Systematic checkpointing of the machine state makes 
restart of execution from a safe state possible upon detection of an error. The
time and energy overhead of checkpointing, however, 
grows with the frequency of checkpointing.  Amortizing this overhead becomes
especially challenging, considering the growth of expected error rates, 
as checkpointing frequency tends to increase with increasing error rates.  Based
on the observation that due to imbalanced technology scaling, recomputing a data
value can be more energy efficient than  
retrieving (i.e., loading) a stored copy,
this paper explores how recomputation of data values (which otherwise would be read from a
checkpoint from memory or secondary storage) can reduce the machine state 
to be checkpointed, and thereby reduce the checkpointing overhead.  
Specifically, the resulting {\em amnesic} checkpointing framework \arch\ 
can reduce the storage overhead by up to 23.91\%; time overhead, by 11.92\%; and
energy overhead, by 12.53\%, respectively, even in a relatively small scale
system.

%% file: checkpoint_intro.tex
\noindent Scalable checkpointing is the key to enable emerging
high-performance computing applications. 
Ready to expand their problem sizes as more hardware resources (e.g., more cores
under weak scaling)
become available, these applications
challenge processing capabilities.
More hardware resources translate into more components subject to errors, which,
along with a higher expected component error rate as an artifact of technology
scaling, results in a higher probability of (system-wide) errors. 
Therefore, proper error detection and recovery becomes a must for successful
completion of any execution.

Systematic (often, periodic) checkpointing of the machine state enables backward
error recovery (BER) upon detection of an error, by rolling back to and
restarting 
execution from a {\em safe} (i.e., error-free and consistent) machine state. 
Energy and time
overhead of checkpointing the machine state, however, grow with the
frequency of checkpointing. The expected increase in error rates makes
amortization of this overhead especially challenging, as a higher probability of
error directly implies more frequent checkpointing.

The overhead of BER spans the overhead of checkpointing and the overhead of
recovery (which entails roll-back + restart).  The time or energy overhead of checkpointing, $o_{chk}$,
applies every time the system generates a checkpoint; the time and energy
overhead of recovery,
$o_{rec}$, every time the execution restarts from the most recent checkpointed
(safe) state after detection of an error.
Depending on the 
interaction among parallel tasks of execution during checkpointing and recovery,  
BER schemes typically form two major classes: {\em coordinated} and {\em
uncoordinated}~\cite{Johnson1988, Lee1990}.
Coordinated schemes enforce tight lock-step coordination (i.e., synchronization)
among all parallel tasks every time the system generates a checkpoint or
triggers recovery, and hence, generally incur a higher overhead.  
Uncoordinated schemes address this overhead by omitting coordination or
confining it only to tasks interacting with each other during computation, which
as a downside complicates the establishment of  
a consistent error-free global state. 

The checkpointing overhead, 
$o_{chk}$  is proportional to the 
time or energy spent on storing the checkpointed state (to memory or secondary
storage), $o_{wr,chk}$, and the
number of checkpoints,
$\#_{chk}$ (which represents a proxy for the checkpointing frequency).
Putting it all together, 
\begin{equation}
o_{chk} =  \#_{chk} \times o_{wr,chk}
\label{eq:chk}
\end{equation} 
applies. The recovery overhead, $o_{rec}$, on the other hand, includes the time
or energy (spent on useful work and) lost since the 
most recent safe checkpoint, $o_{waste}$, and the time or energy spent on
restoring the state 
captured by the 
most recent safe checkpoint, $o_{roll-back}$.  Under an error
probability
of $perr$, which dictates the number of recoveries, the 
recovery overhead becomes:
\begin{equation}
o_{rec} = perr \times (o_{waste} + o_{roll-back})
\label{eq:rec}
\end{equation}

Imbalances in technology scaling render the energy consumption (and latency) of
data storage and communication significantly higher than the energy consumption
(and latency) of actual data generation, i.e., computation~\cite{Kogge, Horow}.
As a result, whenever a data value is needed (i.e., has to be loaded from
memory), re-generating (i.e., {\em recomputing}) the respective value can easily
become more energy-efficient than retrieving the stored copy 
from memory~\cite{amnesiac17}.
During recovery, recomputation of a data value, which otherwise would be read
from a checkpoint, can therefore be less energy hungry and time consuming than
retrieving the respective checkpoint from main memory or secondary storage. This
can further eliminate the need for checkpointing such {\em recomputable} data
values, which would never be retrieved from memory or secondary storage, but
recomputed.
The result is an {\em amnesic} BER framework, \arch, which can opportunistically
omit checkpointing of (recomputable) data values, and thereby can reduce 
the machine state to be checkpointed, by
relying on the ability to recompute the respective data values when needed
during recovery.  

Under recomputation, time or energy spent on storing the checkpointed state,
$o_{wr,chk}$, can decrease
since a (recomputable) subset of the 
updated memory values would be omitted from checkpointing.
This in
turn can decrease $o_{chk}$, even if $\#_{chk}$ remains the same.
However, the recovery overhead $o_{rec}$
now has to incorporate the overhead of recomputation (of the values which were
omitted from checkpointing), $o_{rcmp}$. 
Still, we expect the time or energy spent on restoring the state of the 
most recent safe
checkpoint, $o_{roll-back}$ to decrease,
since the size of checkpoints would simply reduce under recomputation.
Putting it all together,
the recovery overhead under recomputation becomes: 
\begin{equation}
o_{rec,rcmp} = perr \times (o_{waste,rcmp} + o_{roll-back,rcmp} + o_{rcmp})
\label{eq:recRCMP}
\end{equation}
\noindent 
Therefore, for \arch\ to hold recovery overhead at bay,
$o_{rec,rcmp} \leq o_{rec}$ 
should be the case, which implies:
\begin{equation}
o_{roll-back,rcmp} + o_{rcmp} \leq o_{roll-back}
\label{eq:rcmpBE}
\end{equation}

{\em Recomputation} in this case is fundamentally different than classic {\em
replay}: {\em recomputation} refers to the recalculation of a data value to cut any
energy-hungry memory access associated with the respective value. This can be
regarded as restricted {\em replay} of a small backward slice of instructions
just to generate that respective data value.

In this paper, we explore how
\arch\ 
can help reduce the overhead of checkpointing without compromising the overhead
of recovery in
terms of time, energy, and storage.
\arch\ is:

\begin{list}{\labelitemi}{\leftmargin=1em}
\vshrink{0.2}
\itemsep-0.3em 
\item {\em hybrid (hardware/software)}: \arch\ relies on a
	  compiler pass to generate (and embed into the binary) instructions
	  required to recompute the respective data values, which can be excluded
	  from checkpointing. 
	  Under recovery, \arch's runtime scheduler in turn triggers recomputation
	  of these values.    
 \item {\em transparent}: Both, 
   {\em amnesic}
   binary generation and
 triggering recomputation upon recovery are transparent to the application
 developer and user.  
 \item {\em low overhead}: 
  \arch\ trades the data storage and retrieval overhead of checkpointing for the
  overhead of recomputing the respective data values. \arch\ can
  significantly reduce the overhead of checkpointing,    
   while holding
   recomputation-incurred overheads (particularly during recovery) at bay. 
 \item {\em scalable}: Traditional checkpointing and recovery becomes more
   challenging at larger scale. 
   \arch\ can
   effectively reduce the 
   checkpoint size,
 hence, is by construction more scalable.
\vshrink{0.2} 
\end{list}
In the following, we will detail a proof-of-concept \arch\ implementation.
Specifically, Section~\ref{sec:back} provides the background;
Section~\ref{sec:checkpoint_design} discusses \arch\ basics;
Sections~\ref{sec:checkpoint_setup} and~\ref{sec:checkpoint_eval} provide the
evaluation; 
Section~\ref{sec:checkpoint_related} covers the related work; 
and Section~\ref{sec:conc} concludes the paper.

%% file: back.tex

\subsection{Backward Error Recovery (BER)}
\label{sec:checkpoint_ckpt}
\input{checkpoint_ckpt}

\subsection{Data Recomputation for Energy Efficiency}
\label{sec:checkpoint_recomp}
\input{rcmp_short}

%% file: checkpoint_ckpt.tex
\noindent {\bf Checkpointing:}
Checkpointing serves establishment of a safe (i.e., error-free and consistent)
machine state to roll-back to and recover from upon detection of a error,
thereby ensuring forward progress in execution in the presence of errors.  
Without loss of generality, we consider shared memory many-cores featuring directory-based
cache coherence. 
We start our analysis with
global coordinated checkpointing and recovery~\cite{Tamir84,Morin1996,revive,safetyNet}, but provide a sensitivity
study for local coordinated schemes~\cite{Koo1987,Leu1988}, as well.
Under global
checkpointing, 
all cores periodically
cooperate to checkpoint the respective machine state. Specifically, 
at the beginning of each checkpointing period, all cores stop
computation to participate in checkpoint generation.

As a running example (and a relatively lower-overhead baseline for comparison,
not to favor \arch),
we will use a log-based incremental
in-memory checkpointing variant similar to~\cite{rebound,revive,safetyNet},
where 
upon each memory update, a record for the old value goes into a log stored in
memory. This log corresponds to the checkpoint. The log constitutes a record of
values updated only within the time window between two consecutive
checkpointing events, as opposed to the entire machine state. 
Establishing a checkpoint involves writing all dirty cache lines back to memory
and recording (the rest of)
each core's architectural state.
For dirty lines, the memory controller only updates the log with the
corresponding old value, if the update represents the very first modification 
since the last checkpoint.
Thus, similar to~\cite{revive}, a modified cache line gets logged only once
between a pair of consecutive checkpoints. 
The directory controller keeps an additional  
bit per memory line to keep track of 
whether the line has already been logged for the current checkpoint interval. 
The controller sets this bit upon logging the line, and clears it upon
establishing a new checkpoint. 
In the following, we will refer to this bit as \texttt{log}.

In-memory checkpointing, by construction, incurs a lower time and energy overhead
when compared to (more traditional) checkpointing to secondary storage.
In-memory checkpointing may correspond to a stand-alone checkpointing scheme or
represent the first level in a hierarchical checkpointing framework. Our
observations generally apply under both options. 

\begin{figure}[htp]
  \vshrink{0.1}
  \begin{center}
	\includegraphics[width=0.35\textwidth]{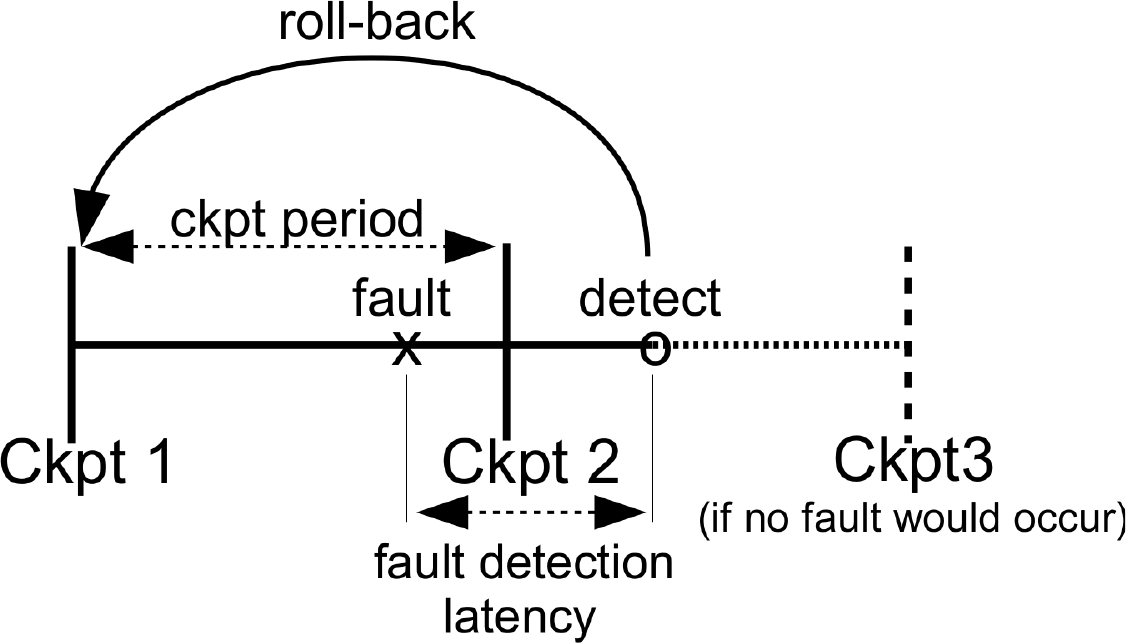}
	\caption{Recovery from an error.}
	\label{fig:ckpt}
  \end{center}
  \vshrink{0.3}
\end{figure}

\noindent {\bf Error Detection and Recovery:}
 
In the following, we assume a fail-stop error model, where 
data memory and checkpoint logs do not suffer from any
errors, similar to~\cite{rebound}.
Various protection mechanisms such as ECC~\cite{Gong2015} 
or memory raiding~\cite{Dell1997} can achieve this.
To detect errors, the system can rely on modular redundancy~\cite{Nomura2011}
or error detection codes (e.g., CRC). 
Error detection is not instantaneous, therefore, a lag between the occurrence of
an error and its detection generally applies, which is 
referred to as error detection latency.
As a consequence, corrupted state may get checkpointed, even if the error
detection latency is no longer than the checkpoint period.
Figure~\ref{fig:ckpt} illustrates an example, where an error occurs right before
{\em Ckpt2} gets taken, and is detected only after {\em Ckpt2} is established,
thereby corrupting the respective checkpointed state. 
In this particular case, 
the time elapsed between establishment of {\em Ckpt2} and the detection of the
error is less than the error detection latency,
hence, there is no guarantee for {\em Ckpt2} to be error-free.
To recover from the error, the system should roll-back to the second most recent
checkpoint at hand, i.e., {\em Ckpt1}, instead of the most recent {\em Ckpt2}.
If the error
detection latency is no longer than the checkpoint period, which applies
throughout
this study, keeping most recent two checkpoints suffices.

%% file: rcmp_short.tex
\noindent 
Imbalances in technology scaling render the energy consumption (and latency) of
data storage and communication significantly higher than the energy consumption
(and latency) of actual data generation, i.e., computation~\cite{Kogge, Horow}.
As a result, whenever a data value is needed (i.e., has to be loaded from
memory), re-generating (i.e., {\em recomputing}) the respective value can easily
become more energy-efficient than retrieving the stored copy 
from memory~\cite{amnesiac17}.
The basic idea behind data recomputation is to eliminate 
memory accesses (be it a read, or a write) by relying on the ability to
recalculate the respective data
values, when needed.  
To this end, the system has to record the 
sequence of
instructions which can produce the respective data values.
As a representative example, the recently proposed Amnesiac
machine~\cite{amnesiac17} details compiler and (micro)architecture support for
opportunistic substitution of memory reads with a sequence of arithmetic/logic
instructions to recompute the data values which would otherwise be retrieved
from the memory hierarchy.  Following Amnesiac's terminology, we will refer to
these sequences of instructions as {\rs}s, each forming a backward slice of
arithmetic/logic instructions.
To perform recomputation along an \rs, its input operands should be available at
the expected time of recomputation.  Not all \rs\  input operands suit
themselves to (re)generation by recomputation, particularly, if input operands
correspond to read-only values residing in memory (e.g., program inputs),  or
register values which are overwritten at the time of recomputation. Amnesiac
refers to such input operands as {\em non-recomputable} inputs, and to make sure
that they are available at the anticipated time of recomputation, stores them in
designated buffers.
To facilitate recomputation, we assume similar hardware-software support as
Amnesiac, with Section~\ref{sec:checkpoint_design} detailing the fundamental
differences.

%% file: checkpoint_design.tex
\noindent 
In this section, we cover the basics and execution semantics of a practical \arch\
implementation under checkpointing, and recovery upon the onset of an error. 

\vshrink{-.1}
\noindent{\bf Impact on Checkpointing:} At the end of each checkpointing
interval, \arch\ identifies and omits the {\em recomputable} subset of data
values (which otherwise would be included in the checkpoint being taken) from
checkpointing.  Thereby, \arch\ can reduce the checkpoint size, which in turn
reduces the $o_{wr,chk}$ component of the checkpointing overhead per
Equation~\ref{eq:chk}, i.e., the time or energy spent on storing the
checkpointed state to memory.
At the extreme, all values
which otherwise would be included in a checkpoint may be {\em recomputable}. If
this is the case, \arch\ would also be able to eliminate a subset of checkpoints
entirely, and thereby reduce the $\#_{chk}$ component of the checkpointing
overhead per Equation~\ref{eq:chk}, i.e., the number of checkpoints. 

\vshrink{-.1}
\noindent{\bf Impact on Recovery:} Upon the onset of an error, the {\em amnesic recovery
handler} triggers the recomputation of any data value which was omitted from the
checkpoint being restored. Such recomputation incurs the overhead captured by
$o_{rcmp}$ in Equation~\ref{eq:recRCMP}, but, at the same time, can cut back on the
time or energy spent on restoring the checkpointed state from memory 
(i.e., $o_{roll-back}$ in Equation~\ref{eq:rec}).

\vshrink{-.1}
\noindent {\bf Overview:} 
\arch\ trades the checkpoint storage and retrieval overhead from memory for 
the overhead of recomputing the respective data values.
Accordingly, any practical \arch\ implementation has
to address:
\begin{list}{\labelitemi}{\leftmargin=1em}
\itemsep-0.3em 
\item how to identify recomputable data values in a checkpoint interval;
\item how to omit recomputable data values from a checkpoint; and
\item how to trigger recomputation of the respective data values during recovery.
\end{list}

\subsection{Amnesic Checkpointing}
\label{sec:rec_enabled_chk}

\noindent We will first cover {\em how to identify recomputable data values
which can be omitted from checkpointing}.

\begin{figure}[htp]
  \begin{center}
	\includegraphics[width=0.22\textwidth]{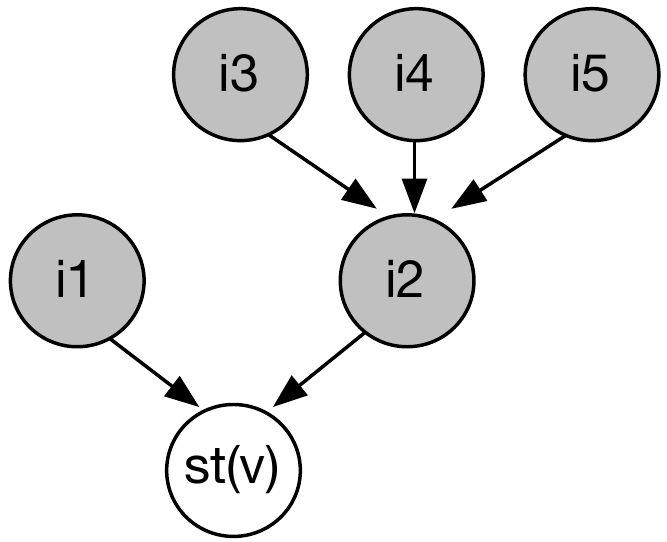}
	\caption{Backward recomputation slice ({\em RSlice}).}
	\label{fig:rslice}
  \end{center}
  \vshrink{0.5}
\end{figure}

\vshrink{-.1}
\noindent {\bf Compiler Support:}
\arch\ relies on a compiler pass to identify recomputable data values,
which can be omitted from checkpointing.  Under incremental in-memory
checkpointing (Section~\ref{sec:checkpoint_ckpt}), only a subset of the store
instructions would trigger checkpointing 
(specifically, only the first updates to the same memory address).  The compiler
pass therefore tracks store instructions, and using data dependency graphs,
extracts backward slices, i.e., sequences of arithmetic/logic instructions which
produce the respective data values to be stored.  Following the terminology
from~\cite{amnesiac17}, we refer to each such backward
slice as an \rs.
Fig.~\ref{fig:rslice} shows an example, where the arrows point
to the direction of dataflow, and each node corresponds to an instruction.
Instructions i3, i4, i5 are producers of the (input operands of) instruction i2; instructions i1 and
i2, of the value {\em v} to be stored by the store instruction {\em st(v)}.
Depending on the specifics of the instruction set architecture (ISA), such backward
slices can take different forms. 

In selecting which {\rs}s to embed into the binary, the compiler has choice. One
option is, using probabilistic analysis, estimating the anticipated cost of
recomputation along each \rs\ when compared to reading, i.e., loading the
respective data value from a checkpoint in memory, and including the {\rs} only
if more cost-effective (where cost can be delay, energy or a combination of
both, without loss of generality).  In this study, we instead take a more greedy
approach of minimal complexity, and consider all {\rs}s which have a lower
number instructions than a preset threshold (which typically remains less than
10, and in Section~\ref{sec:checkpoint_eval} we quantify the impact). The
insight is that the overhead of recomputation along an \rs\ increases with its
number of instructions.  Therefore, capping the instruction count can
effectively hold recomputation overhead under control (as we will further
demonstrate in Section~\ref{sec:checkpoint_rslice_length}).  

The next question is how to embed {\rs}s into the binary, to facilitate
invocation upon recovery. The only critical piece of information is associating
the start address of each \rs\ (i.e., the address of the first
instruction in the backward slice) with the memory address of the respective
data value (which will be regenerated by recomputation along the \rs). Such
memory addresses correspond to the destination memory addresses of the stores,
and
the compiler uses each such store as a proxy in identifying target values for recomputation.
One way to communicate this information to the runtime is introducing a special
instruction to associate these two effective addresses (and enforcing atomic
execution of it with the corresponding store). We will refer to this instruction
as \texttt{ASSOC-ADDR}. 

While the compiler analysis to bake recomputing instructions into the binary
looks similar to the compiler pass in~\cite{amnesiac17}, there is a
fundamental difference: The goal in~\cite{amnesiac17} is swapping each energy-hungry
load with an \rs\ to recompute the respective data value (which otherwise
would be loaded from the memory hierarchy). In this case, the swapped load
instructions are never performed.  
In exploiting recomputation for checkpointing, on the other hand, \arch\ leaves
load instructions intact, and only tracks store instructions to identify data
values which can be omitted from checkpointing. In this case, the corresponding
store instructions are always performed; what is omitted is the inclusion of the
respective (recomputable) data value into the corresponding checkpoint.

\vshrink{-.1}
\noindent {\bf Amnesic Checkpoint Handler:}
Each time an \texttt{ASSOC-ADDR} instruction is encountered, 
{\em amnesic checkpoint handler} records the corresponding 
\texttt{<memory address,RSlice address>} 
association into a dedicated buffer called {\em Address Map},
\texttt{AddrMap}. Next, the handler asks the memory controller to exclude the
corresponding (recomputable) value from the next checkpoint (which is achieved by setting the
dedicated \texttt{log} bit, as explained in Section~\ref{sec:checkpoint_ckpt}).
Eventually, the size of the next checkpoint reduces as more
(recomputable) values are excluded from checkpointing via \texttt{ASSOC-ADDR}
instructions.
Such \texttt{<memory address,RSlice address>} pairs have to remain in \texttt{AddrMap}
as long as the
established checkpoint for the corresponding interval remains in memory, such
that 
upon detection of an error, recomputation along {\rs}s can restore the values
omitted from checkpointing, in coordination with the established checkpoint for
roll-back.
As covered in Section~\ref{sec:checkpoint_ckpt}, under the assumption that the error
detection latency does not exceed the checkpointing period, retaining two most
recent checkpoints suffices. Therefore, \texttt{ASSOC-ADDR} should only record
the mappings for the two most recent checkpoints. 

\subsection{Amnesic Recovery}
\label{section:rec_enabled_rec}

\noindent Upon detection of an error, {\em amnesic recovery handler} orchestrates
roll-back to the most recent 
safe
global
recovery line, by triggering recomputation along {\rs}s for each value excluded
from checkpointing, in coordination with the restoration of the most recent safe
checkpoint.
There is no need for separate bookkeeping for the values
missing from the most recent safe checkpoint, since \texttt{AddrMap} contains
all the necessary information to fire recomputation of these values along the
respective {\rs}s.
After recomputing the missing values and storing them back to their destination
addresses, {\em amnesic recovery handler} restores the remaining states in the
checkpoint, and resumes execution
from this point onward. 

In this study, we confine recomputation to memory values only. Therefore, upon
recomputation of a missing value from the checkpoint, we have to access memory
to store the respective value. Register values are checkpointed, as well, as
part of the architectural state, but are not considered for recomputation. This
is likely to render the proof-of-concept \arch\ implementation conservative, as a
register value would not incur an expensive memory write upon recomputation.
In the end, during recovery, \arch\ can only cut the overhead of retrieving
(i.e., loading) the
checkpointed state from memory (due to the omission of recomputable values from
the checkpoint), which can be easily masked by the overhead of writing such
omitted (memory) values back to memory upon recomputation.

\subsection{Microarchitecture Support}
\label{sec:micro_sum}
\noindent To facilitate amnesic checkpointing, the memory controller 
takes a similar form to~\cite{revive}, and maintains
the \texttt{log} bit to determine if the old value of a given write-back
should be logged (i.e., checkpointed).
For each write-back request, the memory controller has to decide (i) whether
the request would result in the first update to the respective memory line since
the last checkpoint was taken, and (ii) whether the current data value $v$ of
the respective memory line (i.e., the value before the write-back takes place)
can be recomputed. While the memory controller can manage the \texttt{log} bit itself for (i),
it should coordinate with {\em amnesic checkpoint handler} for (ii). 
As explained in Section~\ref{sec:rec_enabled_chk}, upon encountering a recomputable
value, the 
{\em amnesic checkpoint handler}
sends a request to the memory controller to let it know that the
respective value $v$ can be recomputed, and therefore, should be omitted from
checkpointing.  The
memory controller sets the \texttt{log} bit accordingly, when it receives such
requests from the {\em amnesic checkpoint handler}.

The number of (stores corresponding to the) values that can be excluded from
checkpointing depends on the size of \texttt{AddrMap}, specifically, 
on how many {\rs}s 
 \texttt{AddrMap} can keep track of.  Fortunately, we do not need an
excessively large \texttt{AddrMap} to this end: 
Recall that we only need to checkpoint the old values upon the very first
write-backs (to unique addresses) when a new checkpoint is established.
Therefore, the number of {\rs}s is not a function of how many times an address
is updated, but {\em how many unique memory addresses} are updated within a given
checkpoint interval.  Naturally, the latter is bounded by the period of
checkpointing. As the period gets longer, the probability of having a higher
number of unique memory addresses updated increases.  At the same time, as the
period gets longer, the amount of useful work lost upon detection of an error
increases.  The checkpointing period cannot get too long
  to reduce this amount of useful work lost.
  The
  checkpointing period hence puts an upper bound on how many unique {\rs}s we
  should keep track of 
  at runtime.
Finally, to prevent corruption of architectural state during recomputation,
\arch\ relies on a similar renaming scheme as~\cite{amnesiac17}.


\subsection{Putting It All Together}
\label{section:overhead}

\noindent 
\arch\ can reduce the 
number of values 
to be logged for checkpointing, and thereby reduce both the performance and
energy overhead of checkpointing.
\arch\ can also 
reduce the size of each checkpoint, and thereby the storage overhead, by cutting 
the number of values to be checkpointed in each interval.
A
reduction in checkpoint size can easily translate into energy savings, as well as
performance gain, due to the lower number of expensive memory read (during recovery)
and write operations (during checkpointing), respectively.

Recovery upon detection of an error involves
recomputation of missing values from the checkpoint and restoring the rest of
the state using the established checkpoint.  Recomputation along each {\rs}
incurs a performance and energy overhead; however, it is not prohibitive since
the number of instructions in {\rs}s are bounded.  During recovery, \arch\
introduces the extra overhead of recomputation, but at the same time, it reduces
the number of values to be read from the checkpoint in memory for restoration.
The 
benefit of the latter 
may or may not be comparable to the overhead of recomputation.
However, considering the anticipated frequency of checkpointing and recovery, one can argue that recovery is a much less frequent event compared to checkpointing, thus \arch's gain under checkpointing is more
likely to outweigh its potential loss under recovery.

%% file: checkpoint_setup.tex
\noindent To evaluate the impact of {\em amnesic} checkpointing and recovery on execution time and energy, we
experimented with eight benchmarks from the NAS~\cite{nas} suite\footnote{with
  the exception of ep due to
simulation complications}.
We ran these benchmarks with 8-32 threads on a simulated 8-32 core system.
We implemented recomputation, checkpointing, and recovery under \arch\ in
Snipersim~\cite{sniper}.
We extracted energy estimates from McPAT~\cite{mcpat} 
integrated with 
Snipersim.
Table~\ref{table:architecture} summarizes the configuration for the simulated
architecture. 

\begin{table}[htp]
  \centering
  \scalebox{0.95}{
\begin{tabular}{|l||l|l|l|}
\hline \hline
\multicolumn{4}{|l|} {Technology node: 22nm} \\
\hline 
\multicolumn{4}{|l|} {Operating frequency: 1.09 GHz}\\
\multicolumn{4}{|l|} {4-issue, in-order, 8 outstanding ld/st}\\
\hline 
\hline
L1-I (LRU): & \multicolumn{3}{|l|} {32KB, 4-way, 3.66ns}\\
\hline 
L1-D (LRU, WB): & \multicolumn{3}{|l|} {32KB, 8-way, 3.66ns}\\
\hline 
L2 (LRU, WB): & \multicolumn{3}{|l|} {512KB, 8-way, 24.77ns}\\
\hline 
\hline
Main Memory & \multicolumn{3}{|l|} {120ns, 7.6 GB/s/controller}  \\
   & \multicolumn{3}{|l|} {1 mem. contr. per 4-cores } \\
\hline
Network Bandwidth &\multicolumn{3}{|l|} {128 GB/s}\\
\hline \hline
\end{tabular} 
  }
  \vshrink{0.1}
  \caption{Simulated architecture. }
  \label{table:architecture}
\end{table}

We implemented \arch's
compiler pass to embed {\rs}s into the binary
as a Pin~\cite{pin} tool. Recall that Snipersim relies on a Pin-based front-end,
which facilitated seamless integration. We used a predetermined threshold
for \rs\ length:
{\rs}s exceeding threshold are 
excluded from the
binary to prohibit
excessive recomputation overhead along \rs\/s. In Section~\ref{sec:checkpoint_rslice_length}, we will
discuss
the impact of the threshold value on checkpointing overhead.  

We considered the following configurations:
\begin{list}{\labelitemi}{\leftmargin=1em}
\itemsep-0.3em 
\item \nockpt: Error-free execution without any checkpointing or recovery
  support. This baseline does not incur any checkpointing or recovery overhead.
\item \nfnr: Periodic coordinated global checkpointing under error-free
  execution, which incurs no recovery overhead.  Only checkpointing overhead
  becomes visible.
\item \wfnr: Periodic coordinated global checkpointing in the presence of errors,
   such that recovery overhead becomes visible on top of checkpointing overhead.
\item \nfwr: \arch\ incorporated into coordinated global checkpointing, under error-free execution, which incurs no recovery
  overhead. Only checkpointing overhead becomes visible.  
\arch\ can reduce checkpoint size by
omitting data values from checkpointing.
\item \wfwr: \arch\ incorporated into coordinated global checkpointing, in the presence of errors, such that recovery overhead
  becomes visible on top of checkpointing overhead.
  \arch\ can reduce checkpoint size by
  omitting data values, which can be recomputed upon recovery, from checkpointing.
\item \nfnrl: Coordinated local checkpointing under error-free
  execution, which incurs no recovery overhead.  Only checkpointing overhead
  becomes visible.
\item \wfnrl: Coordinated global checkpointing in the presence of errors,
   such that recovery overhead becomes visible on top of checkpointing overhead.
\item \nfwrl: \arch\ incorporated into coordinated local checkpointing, under error-free execution, which incurs no recovery
  overhead. Only checkpointing overhead becomes visible.  

\arch\ can reduce checkpoint size by
omitting data values from checkpointing.
\item \wfwrl: \arch\ incorporated into coordinated local checkpointing, in the presence of errors, such that recovery overhead
  becomes visible on top of checkpointing overhead.
  \arch\ can reduce checkpoint size by
  omitting data values, which can be recomputed upon recovery, from checkpointing.
\end{list}

We adjust the checkpointing frequency to the expected error rates
and the execution times of the applications.
Without loss of generality, we distribute the checkpoint intervals uniformly
over the execution time.
As a result, applications with longer execution times checkpoint more.

%% file: checkpoint_eval.tex
\subsection{Checkpointing Overhead}
\input{checkpoint_ckpt_overhead}

\label{sec:checkpoint_ckpt_overhead}

\subsection{Recovery Overhead }
\input{checkpoint_recovery_overhead}
\label{sec:checkpoint_recovery_overhead}

\subsection{Storage Complexity}
\input{checkpoint_ckpt_reduction}

\label{sec:checkpoint_ckpt_reduction}

\subsection{Coordinated Local Checkpointing}
\input{checkpoint_lc_gc}

\label{sec:checkpoint_lc_gc}

\subsection{Sensitivity Analysis}

\noindent \subsubsection{Impact of RSlice Length on Checkpoint Size}
\input{checkpoint_rslice_length}

\label{sec:checkpoint_rslice_length}

\noindent \subsubsection{Impact of Error Rate}
\input{checkpoint_fault_rate}
\label{sec:checkpoint_fault_rate}

\noindent \subsubsection{Impact of Checkpointing Frequency}
\input{checkpoint_ckpt_freq}

\label{sec:checkpoint_ckpt_freq}

\noindent \subsubsection{Scalability}
\input{checkpoint_thread_count}
\label{sec:checkpoint_thread_count}

%% file: checkpoint_ckpt_overhead.tex
\begin{figure*}[!t]
	\centering
	\includegraphics[width=\textwidth]{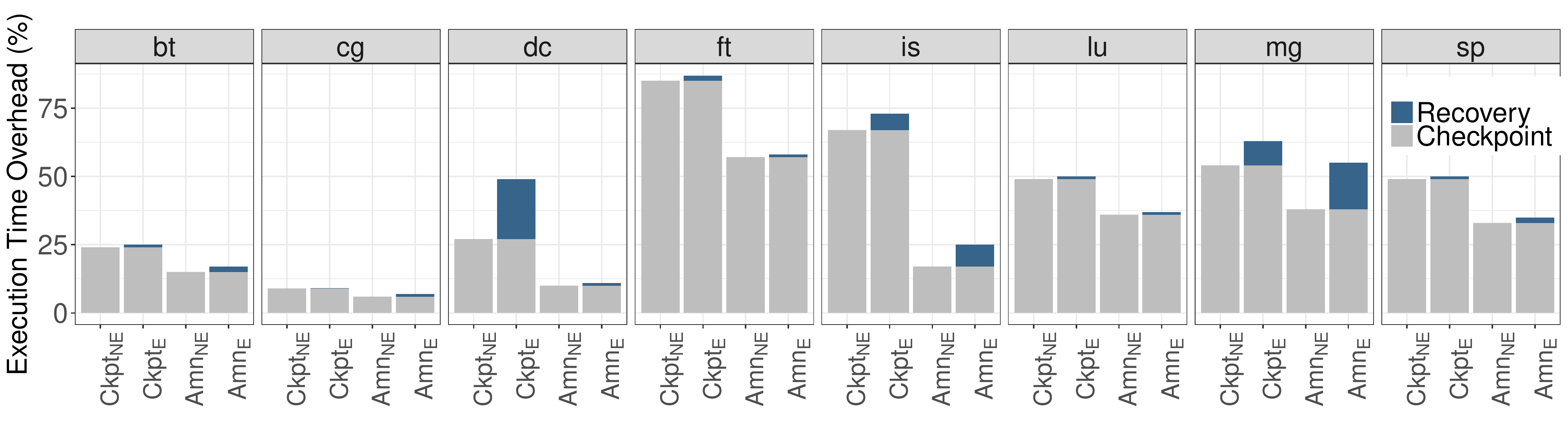}
	\caption{Time overhead of checkpointing and recovery.}
	\label{fig:time_overhead_combined}
\end{figure*}

\noindent 
We start the evaluation with a characterization of the checkpointing overhead
under \arch. For a crisp comparison, we use the configurations from
Section~\ref{sec:checkpoint_setup} under error-free execution, which only incur
the overhead of checkpointing.
Specifically, 
we use \nockpt\ as a baseline for comparison, where
no checkpointing takes place.  Fig.~\ref{fig:time_overhead_combined} shows the execution time overhead of checkpointing and recovery. The first and third columns in each group show the execution time overhead of checkpointing for the evaluated benchmarks under \nfnr\ and \nfwr, respectively.
As expected, \nfnr\ and \nfwr\ perform consistently worse 
than \nockpt\ due to the checkpointing overhead.  
However, via recomputation, 
\nfwr\ is
very effective in reducing the 
\nfnr's time overhead due to checkpointing, by
up to 28.81\% (for
is), and 11.92\%, on average. The smallest reduction is 2.12\% for cg, where 
\nfnr's time overhead is already relatively low. This is because 
cg's checkpoint size per checkpointing interval is relatively small and the
\% of time spent in checkpointing accounts for only $\approx9\%$ of the total execution
time.

\begin{figure*}[!t]
	\centering
	\includegraphics[width=\textwidth]{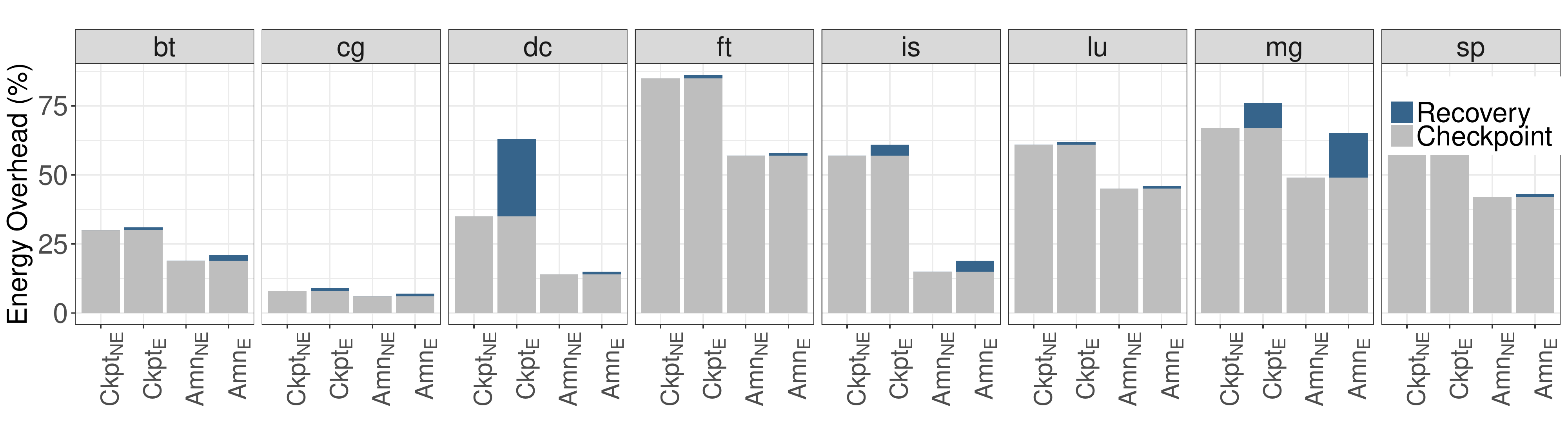}
	\caption{Energy overhead of checkpointing and recovery.}
	\label{fig:energy_overhead_combined}
\end{figure*}

Fig.~\ref{fig:energy_overhead_combined} shows the corresponding energy overhead of
checkpointing and recovery, normalized to \nockpt.
The first and third columns in each group show the energy overhead of checkpointing for the evaluated benchmarks under \nfnr\ and \nfwr, respectively. 
The general trend is similar to the time overhead.  
\nfwr\ reduces the energy overhead of \nfnr\ by up to 26.93\% (for is), and 12.53\%,
on average. Among the benchmarks, {\it is} is very amenable to
recomputation: as the majority of the updated memory values can be 
recomputed
(in case of recovery), \nfwr\ can exclude these from checkpoints, which leads to a
higher reduction in checkpointing overhead w.r.t. \nfnr\/.  The
smallest energy reduction is 1.75\% (for cg), in line with
Fig.~\ref{fig:time_overhead_combined}.

%% file: checkpoint_recovery_overhead.tex
\noindent In Section~\ref{sec:checkpoint_ckpt_overhead}, 
we characterized purely the overhead of checkpointing by assuming error-free
execution where periodic checkpointing still takes place. 
In this section, the goal is quantifying the overhead of recovery, in the
presence of errors. 
Recovery requires the establishment of a globally
consistent state among all cores.  For \wfnr,
this translates into 
each core
rolling back to restore the machine state corresponding to the most recently established
checkpoint. 
This also applies to  \wfwr, but \wfwr\ needs to recompute the data values
omitted from checkpointing, on top.
Such data values have the corresponding \rs\/s baked into the binary.
Therefore, although \wfwr\ 
can reduce the checkpointing overhead, it incurs an extra overhead due to
recomputation during recovery. 
Fig.~\ref{fig:time_overhead_combined}, the second and fourth columns in each group show the execution time overhead of \wfnr\ and \wfwr, respectively (w.r.t \nockpt). Notice that in \wfnr\ and \wfwr, we have an
error during execution.  
As expected, we observe higher time overhead
under \wfnr\ and \wfwr\
than under \nfnr\ and \nfwr,
respectively. 
\wfnr\ and
\wfwr\ both incur the recovery overhead
on top of the checkpointing overhead, as shown in the Fig.~\ref{fig:time_overhead_combined}.
Still, \wfwr\ is very effective in reducing the
time overhead of \wfnr: although \wfwr\ needs to recompute the omitted
values (from checkpointing), thus incurs additional recovery overhead, reduction of checkpointing overhead
(due to the reduced checkpoint size) and reduction of the restore overhead
(again, due to the reduced checkpoint size) outweighs the corresponding overhead of
recomputation. 
As a result, \wfwr\ reduces the time overhead of \wfnr\ by up to 26.68\% (for is), and
12.39\%, on average. The smallest reduction is 1.9\% for cg, in line with our
previous observations. 

The second and fourth columns of each group in Fig.~\ref{fig:energy_overhead_combined} show the percentage of the energy overhead of \wfnr\ and \wfwr\ (w.r.t \nockpt). The energy overhead follows the very same trend as the time overhead.  
\wfwr\ reduces the energy
overhead of \wfnr\ by up to 30\% (for dc), and 13.47\%, on average. The smallest
energy reduction is 1.86\% (for cg). 

\begin{figure}[h]
	\centering
	\includegraphics[width=\columnwidth]{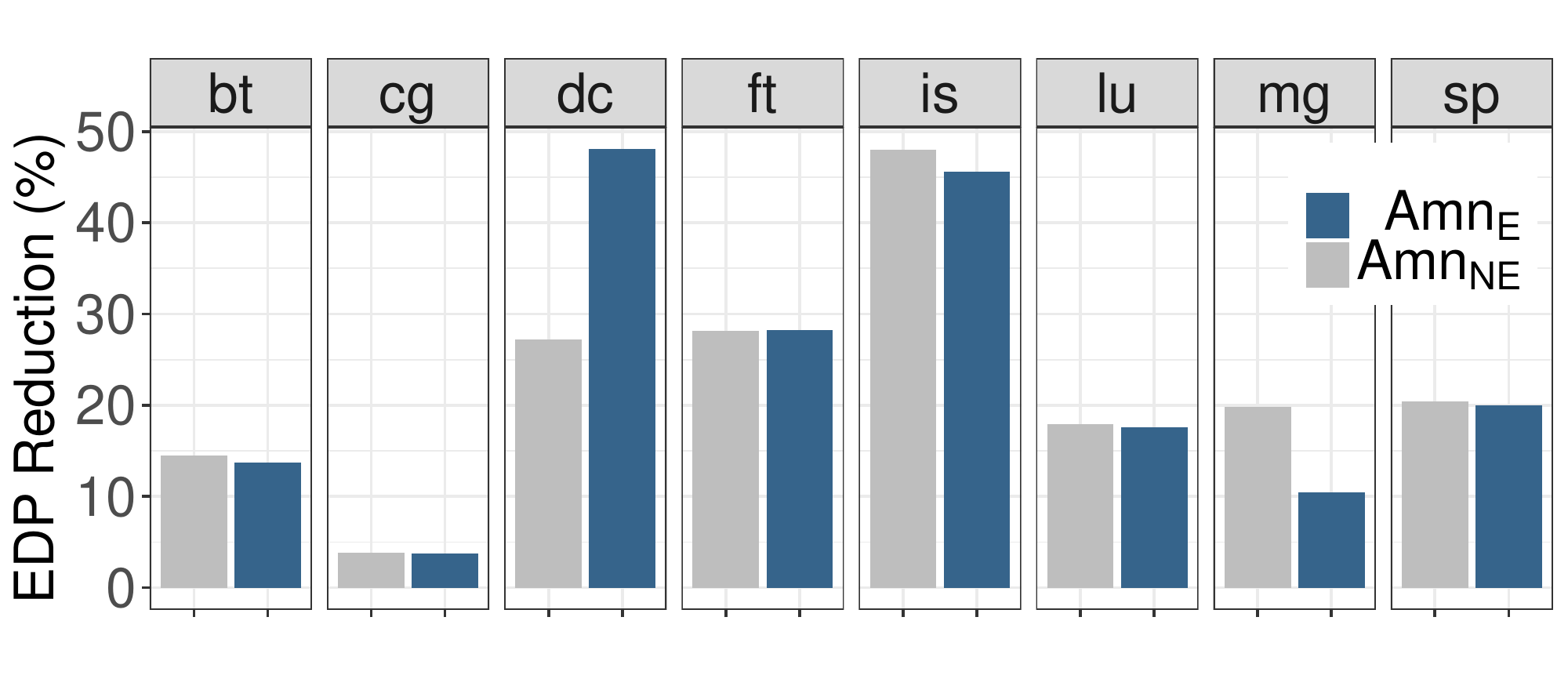}
	\caption{EDP reduction 
		under \nfwr\ and \wfwr\ 
		w.r.t. \nfnr\ and \wfnr\, respectively.}
	\label{fig:edp_reduction_combined}
\end{figure}

Putting it all together, Fig.~\ref{fig:edp_reduction_combined} shows the percentage reduction of energy-delay product
(EDP) of \nfwr\ and \wfwr\ w.r.t. \nfnr\ and \wfnr\, respectively, as a proxy for energy efficiency.
EDP provides a notion of balance between the
time overhead and energy consumption. We observe that \nfwr\ reduces EDP by up to
47.98\% (for is), and 22.47\%, on average, when compared to \nfnr.
Similarly, \wfwr\ reduces EDP by up to
48.07\% (for dc), and 23.41\%, on average, when compared to \wfnr.
Although {\em is} benefits more from \wfwr\ in terms of performance, dc
has a higher energy reduction due to \wfwr, which in turn leads to a higher EDP
reduction.

Overall, we observe that \arch\ can effectively reduce the overhead of checkpointing, as
well as, of recovery.  The effectiveness 
highly depends on the overhead of recomputation along \rs\/s and on how many values can be
omitted from checkpointing. 
We will revisit the impact of \rs\ length on checkpoint size reduction 
in Section~\ref{sec:checkpoint_rslice_length}.

%% file: checkpoint_ckpt_reduction.tex
\noindent
The main benefit of \arch\ stems from the reduction of checkpoint size,
which has two critical implications:
reducing the data size to be (i) {\em moved to} (and {\em retrieved from}); (ii)
{\em stored in} the designated
memory area for checkpointing. 
In addition to (i), (ii) can also reduce the energy consumption, e.g.,  
due to less leakage or refresh in case of DRAM.
At the same time, a reduction in checkpoint sizes can lead to a
reduction in the 
memory footprint of checkpointing, reducing storage complexity. 

\begin{figure}[!t]
\centering
\includegraphics[width=\columnwidth]{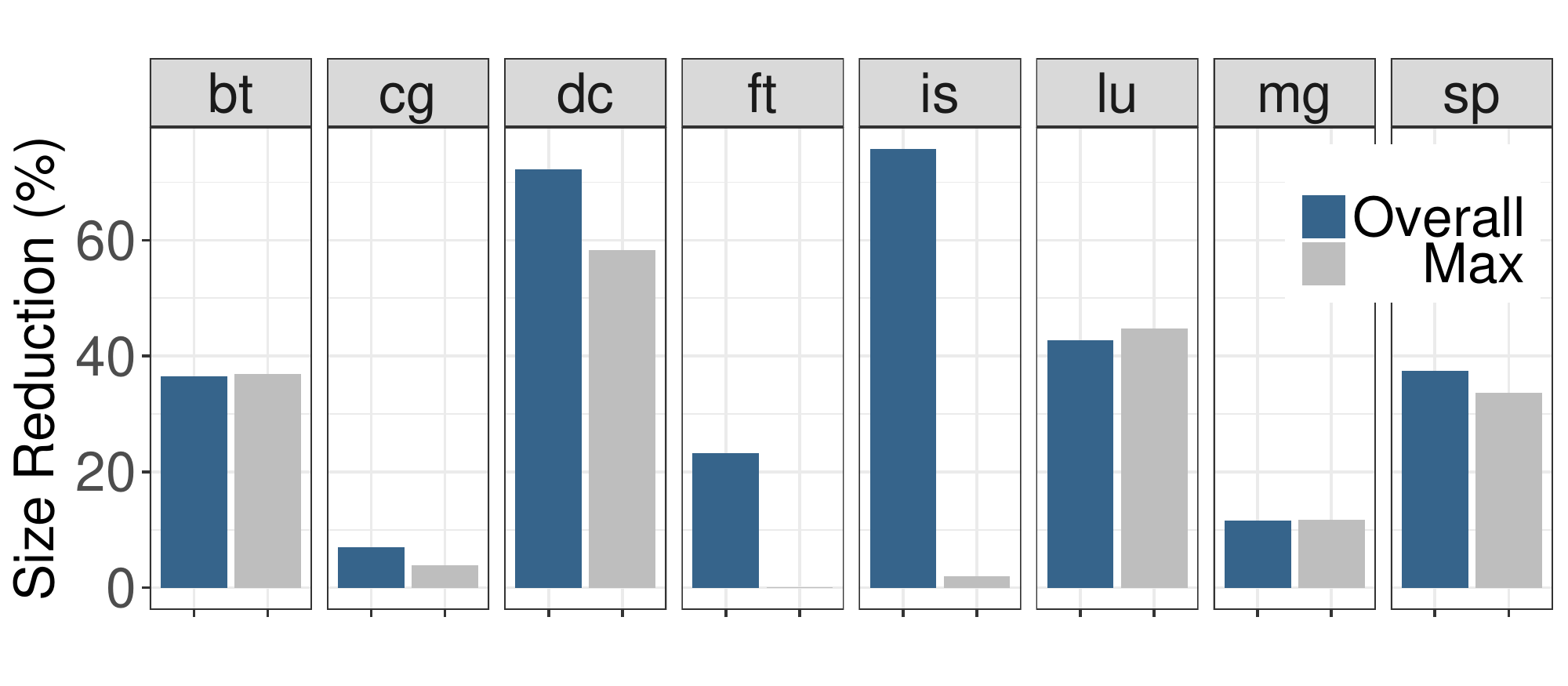}
\caption{\% checkpoint size reduction under \nfwr.}
\label{fig:ckpt_size_reduction_combined}
\end{figure}

The {\em Overall} columns in Fig.~\ref{fig:ckpt_size_reduction_combined} show \% reduction in the 
overall checkpoint size (i.e. total amount of data to be checkpointed) under \nfwr\ w.r.t. to \nfnr. %
Among all benchmarks, {\em is} benefits the most from recomputation, where the
overall checkpoint size reduces by 75.74\% under \nfwr.  On the other hand, cg
is less responsive, and the checkpoint size reduces by 
only 6.99\%.  The average checkpoint size reduction over all benchmarks is
38.31\%.

Recall that, per Section~\ref{sec:checkpoint_ckpt}, 
if the error
detection latency is no longer than the checkpoint period, which applies
throughout
this study, keeping most recent two checkpoints suffices to have ability of recovering the global state (in case of error in execution).
Therefore, the size of the largest checkpoint under \arch\ represents a more
accurate proxy for the anticipated memory footprint reduction than the total
size of all checkpoints (as {\em Overall} columns in Fig.~\ref{fig:ckpt_size_reduction_combined} capture). 
The {\em Max} columns
in Fig.~\ref{fig:ckpt_size_reduction_combined}, hence show
\% reduction in the size of the {\em largest} checkpoint under \nfwr\ w.r.t. to \nfnr.
If there is no value that can be recomputed within the largest
checkpoint, \arch\ cannot reduce the footprint size (although it
may still reduce the the total size of all checkpoints in an application).
Fig.~\ref{fig:ckpt_size_reduction_combined}
reveals such a case: 
{\em is} has
very limited 
{\em Max} reduction (2.04\%) under \nfwr; but the highest {\em Overall}
reduction.
For the rest of the benchmarks,
{dc} shows the largest reduction in {\em Max} of
58.3\%; and {ft},
the smallest 
of 0.05\%.  For {ft}, \arch\ practically cannot
reduce the 
size of largest checkpoint (as the {\em Max} column reveals), but  
the total checkpoint size can still reduce 
by 23.27\% (as the {\em Overall} column reveals).

As explained in Section~\ref{sec:checkpoint_setup}, \nfnr\ and  \nfwr\ exclude
recovery due to error-free execution, hence cleanly capture the overhead, and
particularly size implications of checkpointing.
That said, the corresponding reductions under \wfwr\ would be exactly the same
as under \nfwr, since the presence of errors does not change the set of values
that can be omitted from checkpointing.

%% file: checkpoint_lc_gc.tex
\begin{figure*}[!t]
	\centering
	\includegraphics[width=\textwidth]{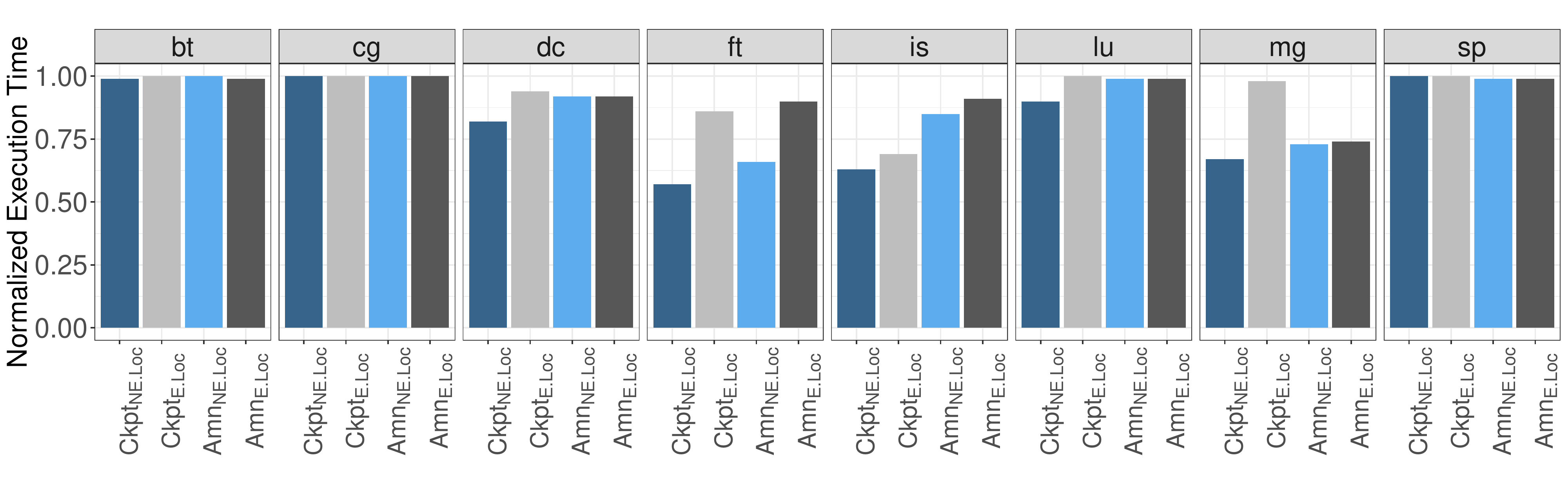}
	\caption{Normalized execution time of \nfnrl, \wfnrl, \nfwrl\ and \wfwrl.}
	\label{fig:lc_vs_gc_combined_time}
\end{figure*}

\noindent In our discussion 
so far we covered coordinated global
checkpointing.
As explained in Section~\ref{sec:checkpoint_ckpt}, a viable alternative is 
coordinated local
checkpointing~\cite{banatre1996,safetyNet}, which  
does not force all cores to participate in
checkpointing: only cores that have been communicating in a given checkpoint interval
checkpoint and rollback (in case of an error)
together.
Coordinated local checkpointing is generally more scalable as the overhead of
checkpointing and recovery evolves with the number of {\em communicating} cores
(as opposed to {\em all} cores under coordinated global checkpointing).
Identifying communicating cores in a checkpointing interval, however, 
necessitates a mechanism to track
inter-core data dependencies,
which usually translates into continuous and dynamic monitoring and recording of
inter-core interactions that may challenge scalability. 
We next investigate recomputation-enabled coordinated local checkpointing. In
the following, we use the global coordinated checkpointing correspondent for
each configuration as a baseline for normalization.

Fig.~\ref{fig:lc_vs_gc_combined_time} shows the normalized execution time
under coordinated local
checkpointing, specifically, \nfnrl, \wfnrl, \nfwrl\ and \wfwrl\ w.r.t. their global checkpointing counterparts (i.e. \nfnr, \wfnr, \nfwr\ and \wfwr, respectively).
We observe that coordinated local checkpointing results in a lower time overhead
for \nfnrl\ 
as indicated by a
y-intercept $<$ 1
for the majority of the benchmarks.
The lower overhead
is due to the lower 
number of cores
checkpointing together. However, this is not the case for 
{\em bt}, {\em cg} and {\em sp}, 
where practically all cores communicate with one
another each checkpointing interval.
For the rest of the benchmarks the time overhead of \nfnrl\
reduces by up to $\approx$42\% for ft, 17\% for
dc, 36\% for is, 32\% for mg, and 10\% for lu w.r.t. \nfnr.

\arch\ incorporated into
coordinated local checkpointing
remains as effective as 
in global checkpointing. For all the benchmarks,
the checkpointing (time) overhead under \nfwrl\ 
remains below (or at most the same as) the overhead under the global
checkpointing correspondent \nfwr.
The reductions under \nfwrl\ are not as pronounced as under \nfnrl,
mainly because the potential
for recomputation does not change considerably under local schemes w.r.t global. 

Specifically, bt, cg, lu, and sp do not observe any sizable reduction
($\approx \leq 1\%$) of the time overhead under \nfwrl\
w.r.t. the global checkpointing
counterpart \nfwr.
For the rest of the benchmarks, the time overhead of
\nfwrl\ reduces by up to $\approx$8\% for dc,
33\% for ft, 15\% for is, and 26\% for mg w.r.t. the global checkpointing
counterpart \nfwr.

We observe similar trends for \wfnrl\ and \wfwrl.
One difference is that the gap in the time overhead w.r.t. to the global
checkpointing counterparts shrinks.
We do not observe any sizable reduction
in the time overhead of {\em bt}, {\em cg}, {\em lu} and {\em sp} under \wfnrl.
For the rest of the benchmarks the performance
overhead of \wfnrl\ 
reduces by up to
$\approx$14\% for {\em ft}, 6\% for {\em dc}, 31\% for {\em is}, and 2\% for {\em mg} 
w.r.t. the global checkpointing
counterpart \wfnr.
On the other hand, the time overhead of \wfwrl\
reduces up to $\approx$8\% for {\em dc}, 10\% for
{\em ft}, 9\% for {\em is}, and 26\% for {\em mg} w.r.t. the global checkpointing
counterpart \wfwr.

The reduction of execution time overhead under coordinated local checkpointing is followed by the EDP reduction. EDP reduces under \nfnrl\ 
by up to
35.68\% for {\em dc}, 
67.15\% for {\em ft},
58.26\% for {\em is},
19.99\% for {\em lu},
and 
57.92\% for {\em mg}
w.r.t. the global checkpointing
counterpart \nfnr.
On the other hand, EDP reduces under \nfwrl\ by up to
15.85\% for {\em dc}, 
55.68\% for {\em ft},
26.24\% for {\em is},
and 
49.75\% for {\em mg}
w.r.t. \nfwr.
Similarly, EDP reduces under \wfnrl\ by up to
18.33\% for {\em dc}, 
33.24\% for {\em ft},
51.46\% for {\em is},
and 
11.29\% for {\em mg}
w.r.t. the global checkpointing
counterpart \wfnr.
On the other hand, EDP reduces under \wfwrl\ by up to
15.80\% for {\em dc}, 
23.81\% for {\em ft},
17.99\% for {\em is},
and 
47.32\% for {\em mg}
w.r.t. \wfwr.

Based on this outcome,
we can conclude that
recomputation-enabled checkpointing and recovery incorporated into coordinated local
checkpointing is at least as effective as 
its global checkpointing
counterpart.

%% file: checkpoint_rslice_length.tex
\begin{figure*}[!t]
	\centering
	\includegraphics[width=0.8\textwidth]{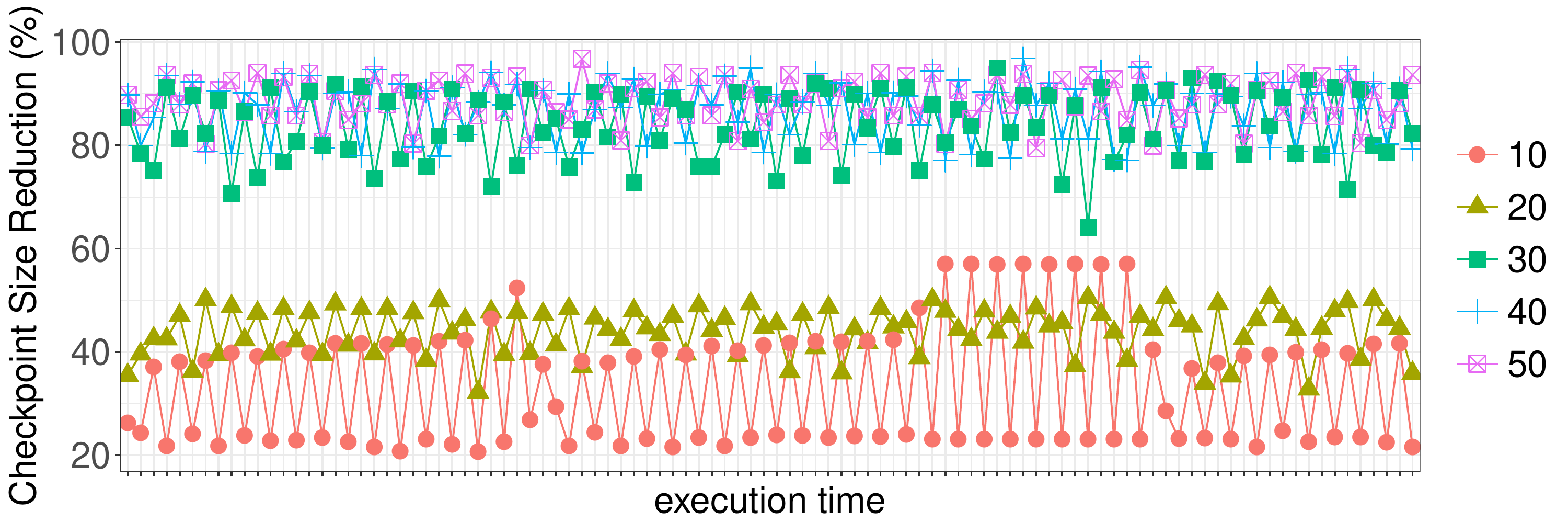}
	\caption{Impact of \rs\ length on checkpoint size over time for bt.}
	\label{fig:bt_ckpt_size}
\end{figure*}
\noindent
\rs\ length (in terms of instructions)
dictates 
the overhead of recomputation. Longer \rs\/s incur a higher
recomputation overhead.  
The overhead of recomputation is invisible under error-free execution, as
recomputation 
may be necessary only during recovery upon detection of an error. 
Throughout the evaluation, we used a threshold of 10 instructions (except {\em
is}, where threshold is 5) to identify the {\rs}s to be embedded into the binary.  

A higher threshold usually translates into 
being able to include more {\rs}s into the binary, and therefore, a higher
likelihood for any value to find a corresponding \rs\ in the binary (and thereby
to get omitted from checkpointing).
As a result, the checkpoint sizes tend to reduce.

Table~\ref{table:rslice_length_threshold} shows
the impact of \rs\ length on the overall 
checkpoint size under \nfwr\. 
As an example, fot {\em bt}, we observe that the total checkpoint size reduces by up to 89.91\%
when the threshold for \rs\ length is allowed to grow up to 50 instructions,  and 36.54\% when the
threshold for \rs\ length remains less than or equal to 10.  
Threshold is a critical design parameter which dictates the overhead of
recomputation
(during recovery in case of an error), and the storage complexity of the
microarchitectural support for \arch\ (as larger buffers are necessary to keep
track of larger {\rs}s).

\begin{table}[!t]
	\centering
	\scalebox{1}{
		\begin{tabular}{|l||l|l|l|l|l|}
			\hline \hline
			\multirow{3}{*}{Benchmark} & \multicolumn{5}{|c|} {Checkpoint Size Reduction (\%)}\\\cline{2-6}
			 & \multicolumn{5}{|c|} {Threshold}\\\cline{2-6}
			 & \textbf{10} & 20 & 30 & 40 & 50 \\
			\hline
			\hline
			bt & 36.54 & 45.14 & 85.36 & 88.36 & 89.91 \\
			\hline
			cg & 6.99 & 67.06 & 89.71 & 89.82 & 89.82 \\
			\hline
			ft & 23.27 & 70.65 & 88.45 & 99.53 & 99.70 \\
			\hline
			ft & 23.27 & 70.65 & 88.45 & 99.53 & 99.70 \\
			\hline
			is\tablefootnote{75.74\% for threshold of 5. Not shown in Table to keep it simple.} & 97.39 & 97.42 & 99.54 & 99.54 & 99.54 \\
			\hline
			lu & 42.69 & 46.65 & 64.43 & 74.69 & 81.11 \\
			\hline
			mg & 11.58 & 19.65 & 87.96 & 90.34 & 90.22 \\
			\hline
			sp & 37.43 & 47.93 & 71.83 & 93.83 & 96.08 \\
	
			\hline \hline
		\end{tabular} 
	}
	\vshrink{0.1}
	\caption{Total checkpoint size reduction as a function of \rs\ Length. }
	\label{table:rslice_length_threshold}
\end{table}

At the same time, 
data values that have the corresponding \rs\/s baked into the binary (and hence
are recomputable) are not necessarily 
uniformly distributed over the checkpoint intervals. Therefore, for each
checkpoint interval, the impact of recomputation may vary (if recomputation is
possible at all).
Fig.~\ref{fig:bt_ckpt_size} shows this effect for {\em bt}, by capturing how \%
reduction in checkpoint size changes over the execution time, considering
different threshold values.  
We observe that \nfwr\ reduces checkpoint size more in certain checkpoint intervals
when compared to others. Such temporal variation points to more optimization
opportunities for \arch: 
for example, instead of checkpointing periodically, adjusting the time to
checkpoint to exploit more recomputation opportunities. 
We leave the exploration of this to future work.

%% file: checkpoint_fault_rate.tex
\noindent The expected (system-wide) error rate ({\em perr}) dictates the
rollback and recovery overhead, as captured by
Equations~\ref{eq:rec} and~\ref{eq:recRCMP}.
Our discussion so far characterized the recovery overhead under \wfnr\ and \wfwr\
assuming a single error within the course of execution. In this section we
expand this analysis to execution under more frequent onset of errors.

\begin{figure*}[!t]
\centering
\includegraphics[width=\textwidth]{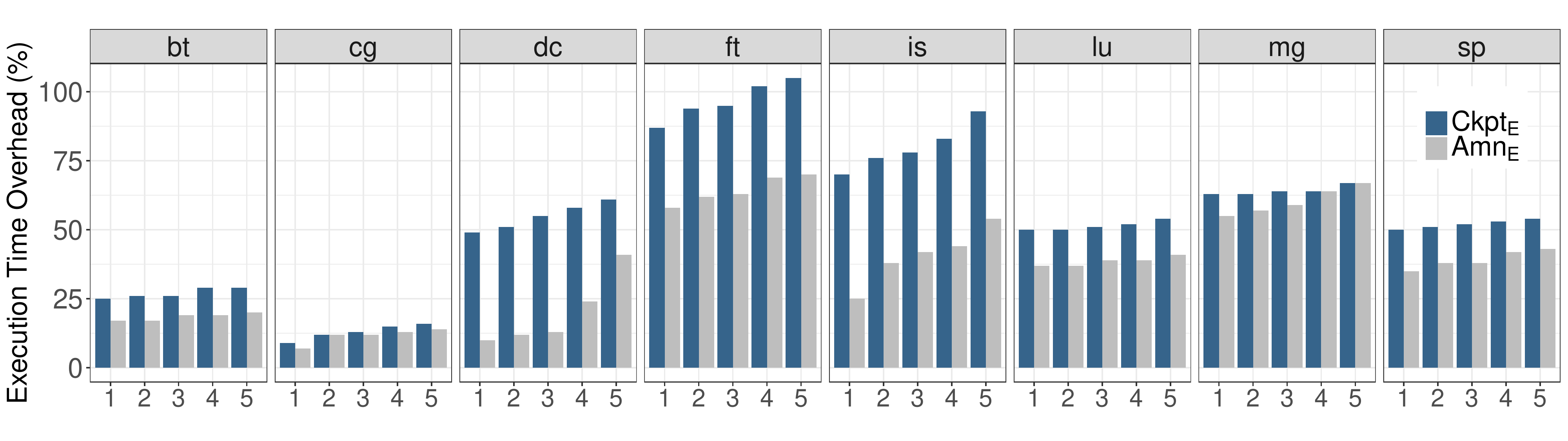}
\caption{Time overhead of \wfnr\ and \wfwr\ considering different error rates.}
\label{fig:error_rate_combined_time}
\end{figure*}

With increasing error rates, the expected number of errors within the course of
execution increases, which in turn increases the recovery overhead due to more
frequent recoveries within the course of execution. 
Fig.~\ref{fig:error_rate_combined_time} shows the \% execution time overhead of
\wfnr\ and \wfwr\ w.r.t. \nockpt, considering different numbers of (up to 5) errors
within the course of execution. We assume that the errors in each case are
uniformly distributed over the execution time.
Not surprisingly, the execution time overhead increases with increasing number of errors. 
Some benchmarks experience very high time overhead as the error rate increases.
This is mainly because the execution time under \nockpt\ is relatively small
such that the overhead of rollback and recovery becomes proportionally higher.
Among the benchmarks, {\em ft} suffers the most as its per recovery overhead is
relatively high.

While the execution time overhead patterns are very similar for \wfnr\ and \wfwr\, the overheads are lower in \wfwr\, since
overall recovery overhead (including restoring the checkpointed values and
recomputing missing values on top) is considerably low in \wfwr.
Specifically, the time
overhead reduces by up to 
26.68\% (for is) for a single error, 
25.35\% (for dc) for two errors,
26.87\% (for dc) for three errors,
21.58\% (for dc) for four errors,
and 19.92\% (for is) for five errors, respectively, 
in \wfwr\ w.r.t. \wfnr. On average, execution time overhead reduction ranges from
$\approx$9\% up to 12\% for different error rates under \wfwr. 

EDP also increases with increasing error rates. 
The general trend is similar to the time overhead, but more pronounced.
Under \wfwr\, EDP reduces by up to
48.07\% (for is) for a single error, 
47.77\% (for dc) for two errors,
50.04\% (for dc) for three errors,
42.99\% (for dc) for four errors,
34.99\% (for is) for five errors.
On average, EDP reduction ranges from $\approx$18\% up to 24\% for different error rates under \wfwr. 

%% file: checkpoint_ckpt_freq.tex
\begin{figure*}[!t]
	\centering
	\includegraphics[width=\textwidth]{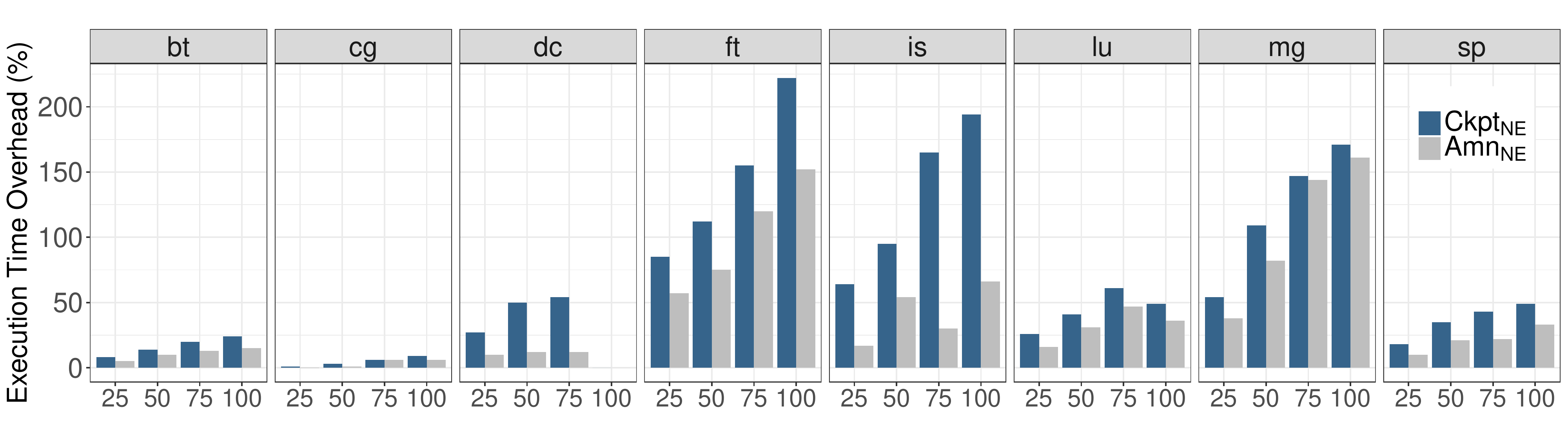}
	\caption{Time overhead of \nfnr\ and \nfwr\ considering
		different numbers of checkpoints.}
	\label{fig:ckpt_freq_combined_time}
\end{figure*}
\noindent
As captured by Equation~\ref{eq:chk}, the time or energy overhead of
checkpointing is a function of the frequency of checkpointing,
as well as the amount of machine state being updated during each checkpointing
interval.
In Section~\ref{sec:checkpoint_fault_rate}, we evaluated the impact of the error
rate on recovery overhead under a fixed checkpointing frequency. In this
section, we evaluate the impact of the checkpointing frequency on checkpointing
overhead under a fixed error rate.
To do so, we vary the checkpointing frequency for each benchmark to yield
25, 50, 75 and 100 checkpoints within the course of execution.  These
checkpoints are uniformly distributed over the execution time.

Fig.~\ref{fig:ckpt_freq_combined_time} shows the execution time overhead of \nfnr\ and \nfwr\  (w.r.t. \nockpt), considering different number of checkpoints. 
Naturally, the time overhead of checkpointing increases with the number of
checkpoints. 
Among all the benchmarks, {\em ft} experiences the largest
time overhead.

The general trend for \nfwr\ is very similar to \nfnr, however, \nfwr\
considerably reduces the time overhead of checkpointing. An interesting point in
Fig.~\ref{fig:ckpt_freq_combined_time} is the lower overhead of 75-checkpointed
runs when compared to 50-checkpointed. 
Although it seems unintuitive at 
first,
there is catch:  when we change
the checkpointing frequency, the start time of each checkpoint interval becomes
different (since we uniformly distribute the checkpoints over the execution time).  The ability of
recomputation to reduce the checkpoint size (and thereby the checkpoint overhead) depends on
whether the corresponding {\rs}s in a given checkpoint interval exist (i.e.,
were baked into the binary). 
If the checkpoints fall into the intervals of execution with a small number of 
recomputable values, \arch\ cannot reduce the checkpointing overhead
significantly. 
Such a corner case is {\em is}, where the 50-checkpointed run has very limited
\rs\ coverage w.r.t. the 75-checkpointed. As the data size 
that can be recomputed (i.e., 
excluded from checkpointing) is smaller, the time overhead is higher for the
50-checkpointed run.
The time overhead 
reduces by up to 
28.81\% (for is) for 25; 
25.3\% (for dc) for 50;
50.86\% (for is) for 75;
and 
43.52\% (for is) for 100 checkpoints
in \nfwr\ w.r.t. \nfnr. On average, the time overhead reduction ranges from
$\approx$10\% up to 14\% for different checkpoint counts in \nfwr.

A similar trend holds for EDP. 
\nfwr\ reduces the EDP (w.r.t. \nfnr) by up to
47.98\% (for is) for 25; 
47.74\% (for dc) for 50;
74.19\% (for is) for 75;
and 
63.45\% (for is) for 100 checkpoints, respectively. On average, EDP reduction
ranges from $\approx$20\% up to 26\% for different checkpoint counts under \nfwr.

%% file: checkpoint_thread_count.tex
\noindent 
The number of threads involved in execution affect the overhead of checkpointing, 
due to both an increase in the cost of coordination (among threads) and a
potential increase in the machine state to be checkpointed. 
As a consequence, the memory bandwidth requirement tends to increase, as well.
We next look into the scalability of \arch\ with increasing thread count. We
experiment with
8-, 16-, and 32-threaded executions where each thread is pinned to a separate core.

We observe that the 
checkpointing overhead always exceeds
9\% for any thread count. On average, the checkpointing overhead is
$\approx$ 45\%, 55\%, and 60\% for 8-, 16-, and 32-threaded executions,
respectively, under \nfnr.
We also observe that \nfwr\ can  
reduce the checkpointing overhead by
up to 28.81\% (for is), 17.78\% (for is), and 19.12\% (for mg).
when running with
8-, 16-, and 32-threads, respectively.
Average reduction is $\approx$12\% for 8-threaded, and $\approx$11\%
for 16- and 32-threaded executions.

The corresponding EDP reduction under \nfwr\ reaches up to 47.98\% (for
is), 31.81\% (for dc), and 33.8\% (for mg) when running with 8-, 16-, and
32-threads, respectively.  Average EDP reduction under \nfwr\  becomes
$\approx$22\%, 21\% and 20\% for 8-, 16-, and 32-threaded executions. 
The corresponding reductions under \wfwr\ closely follow the trends \nfwr.

%% file: checkpoint_related.tex
\noindent 
Checkpointing and recovery solutions are extensively studied over the decades.
The proposed solutions can be categorized into software-based or hardware-based
checkpointing; and application or system level checkpointing.  Software-based
proposals use periodic barriers to perform system-level~\cite{tick},
application-level~\cite{Bronevetsky2004}, or hybrid checkpoints~\cite{C3}. 

Hardware proposals~\cite{rebound,revive,safetyNet} reduce the checkpoint and
restart penalties, but can increase hardware complexity.  For example, in
Rebound~\cite{rebound} when a core is checkpointing, the L2 controller writes
dirty lines back to main memory while keeping clean copies in L2, and the memory
controller logs the old values of the updated memory addresses. In addition,
between checkpoint times, when a dirty cache line is written back to memory, the
memory controller has to log the old value, as well. This is done for the first
write-back and consecutive writes to the same memory address can be excluded from
being logged.  SafetyNet~\cite{safetyNet}, on the other hand,  explicitly
checkpoints the register file, and incrementally checkpoints the memory state by
logging the old values.

Compiler-assisted checkpointing~\cite{Bronevetsky2008} improves the performance
of automated checkpointing by presenting a compiler analysis for incremental
checkpointing, aiming to reduce checkpoint size.  In incremental checkpointing,
memory updates are monitored and are omitted from checkpointing if a particular
memory location has not been modified between two adjacent checkpoints.  This
mechanism reduces the amount of data to be checkpointed, and is widely used in
many checkpointing schemes. We also employ incremental checkpointing in our
analysis. In~\cite{Bronevetsky2008}, instead of using runtime mechanisms (such
as exploiting cache coherency protocol to identify updates memory locations),
they rely on compiler analysis to track the memory updates that can be excluded
from checkpoints.  To facilitate the compiler analysis, the source code should
be manually annotated, indicating the starting point of each checkpoint.
However, it has limited applicability in practice, since it may not be always
feasible to obtain and/or annotate the source code. 

A relevant work presented in~\cite{idem}, introduces the notion of idempotent
execution 
that does not need explicit
checkpoints to recover from errors. 
Instead, in case of an error,
re-executing the idempotent region suffices for recovery.
Such idempotent regions are constructed by the compiler. As the name suggests,
idempotent regions regenerate the same output regardless of how many times they
are executed with the given program state.  In comparison to \arch,  idempotent execution has
limited flexibility.  Generally, idempotent regions are large, and therefore
incur high overhead during recovery, while we employ fine-grained data
recomputation (along a short separate \rs\ for each value), and each \rs\ contains only
the necessary instructions to generate a single value.
Identifying
idempotent regions is also a daunting task, and it may not be easy to find 
fine-grained idempotent regions for a large class of applications.
\rs\/s provide more
flexibility on values to be checkpointed and be recomputed in this regard.

A recent work demonstrates the applicability of recomputation to loop-based code~\cite{Elnawawy2017} to reduce the checkpointing overhead. 
Similar to our approach, they try to reduce the checkpoint size by logging enough state to enable recomputation in case of error in execution. When error occurs, they determine which parts of the computation were not completed and they eventually recompute them by reexecuting the corresponding loop iterations. Although, it is very similar to our approach in spirit, their approach is more restricted to loop-based code, whereas our approach can target arbitrary data as long as its corresponding \rs\ exist. 

Similar to~\cite{Elnawawy2017}, the authors of~\cite{Sloan2013} exploit the regularity of workloads, such as matrix-vector multiplication and iterative linear solver to reduce the performance overhead of checkpointing by relying on partial recomputation.
Their fundamental observation is that although error occurs in computation, 
most of the results are still correct for those types of workloads. So, instead of simply rolling back
and repeating the entire segment
of computation, they employ algorithmic
error localization and partial recomputation to efficiently correct
the erroneous results.

In~\cite{Sayed2014}, authors explore energy concerns for checkpointing and evaluate a wide-range checkpointing policies to understand their respective energy, performance and I/O tradeoffs. They provide detailed insights into
the energy overhead, as well as the performance impact, associated with different checkpointing policies.

%% file: conc.tex
\noindent In the presence of errors, systematic checkpointing of the machine state makes recovery of 
execution from a safe state possible.  
The performance and energy overhead, however, can
become overwhelming with increasing frequency of checkpointing and recovery, as
dictated by the 
growth in the frequency of anticipated errors. 
In this paper, we discuss how recomputation of data values 
which otherwise would be read from a checkpoint (from main memory or secondary
storage)
can help reduce
these overheads.
We observe that recomputation can reduce the memory footprint 
by up to 23.91\%, which 
is accompanied by a reduction in time, energy and EDP overhead by
up to 11.92\%, 12.53\%, and 23.41\%, respectively, even considering a relatively
small-scale system.
We expect the reduction to become much higher and more visible in larger scale
systems, where checkpointing overhead becomes more prominent.